\documentclass[a4paper,twoside]{article}
\pdfoutput=1 

\usepackage{epsfig}
\usepackage{hyperref}
\usepackage{subcaption}
\usepackage{calc}
\usepackage{amssymb}
\usepackage{amstext}
\usepackage{amsmath}
\usepackage{amsthm}
\usepackage{multicol}
\usepackage{pslatex}
\usepackage{comment}
\usepackage{apalike}
\usepackage[bottom]{footmisc}
\usepackage{enumitem}
\usepackage{pifont}
\usepackage{algorithm}
\usepackage{color}
\usepackage{authblk}
\usepackage{algpseudocode}
\newtheorem{definition}{Definition}
\newtheorem{property}{Property}

\theoremstyle{definition}
\newtheorem*{example}{Thread Example}
\begin{document}

\title{An Easy-to-use and Robust Approach for the Differentially Private De-Identification of Clinical Textual Documents}

\newcommand\JFC[1]{\textcolor{green}{#1}}
\newcommand\DL[1]{\textcolor{blue}{#1}}
\newcommand\TODO[1]{\textcolor{red}{#1}}

\newcommand\YT[1]{\textcolor{red}{#1}}

\author{Yakini Tchouka, Jean-François Couchot and David Laiymani}
\affil{{Femto-ST Institute}, {Univ. Bourg. Franche-Comt\'e, CNRS},{France}}


\maketitle
\abstract{Unstructured textual data is at the heart of healthcare systems. For obvious privacy reasons, these documents are not accessible to researchers as long as they contain personally identifiable information. One way to share this data while respecting the legislative framework (notably GDPR or HIPAA) is, within the medical structures,
to de-identify it, \textit{i.e.} to detect the personal information of a person
through a Named Entity Recognition (NER) system and then replacing it to make it very difficult to associate the document with the person. 
The challenge is having reliable NER and substitution tools without compromising confidentiality and consistency in the document. 
Most of the conducted research focuses on English medical documents with coarse substitutions by not benefiting from advances in privacy. 
This paper shows how an efficient and differentially private de-identification approach can be achieved by strengthening the less robust de-identification method and by adapting state-of-the-art differentially private mechanisms for substitution purposes.
The result is an approach for de-identifying clinical documents in French language, but also generalizable to other languages and whose robustness is mathematically proven.}

\section{{INTRODUCTION}}
\label{sec:introduction}
Unstructured textual data is at the heart of healthcare systems.
The details included in these documents allow us to clearly and precisely describe patients' diseases and medical procedures, and to efficiently manage and study their pathologies.
These textual documents can be analyzed by Artificial Intelligence, given the impressive advances in Natural Language Processing techniques in recent years~\cite{kersloot2020natural,velupillai2018using}.

However, on the one hand, these AI-based technologies are currently only accessible to computer researchers and not to medical staff, who have access to medical data.  
On the other hand and for obvious privacy reasons, these medical documents are not accessible to researchers as long as they contain personally identifiable information. 
Medical managers are therefore faced with a familiar dilemma: should they share this medical data and compromise privacy and medical secrecy to allow patients to benefit from the latest medical advances available thanks to the artificial intelligence implemented on this data?   

The GDPR does, however, allow researchers to work on this type of data, provided that it has been anonymized beforehand~\cite[Recital 26]{gdpr}. GPDR is the European legal framework, on the US side we have the Health Insurance Portability and Accountability Act (HIPAA)~\cite{cohen2018hipaa} which defines 18 categories of so-called personal information (PHI) that must be removed from a medical document before it can be shared.
To comply with this legal framework, it is therefore sufficient for medical authorities to provide researchers with de-identified documents. Such a document is a document where the medical information is present but where all 
personal data (names, dates, locations, for example) have been modified to make any identification very difficult.  
 
Practically, this can be done in two steps.
De-identification consists of first, applying a Named Entity Recognition (NER) task revealing words that would allow the document to be re-associated with a particular person. Then, these entities are replaced with alternative words making it very difficult to associate the document with its patient while preserving the utility of the document. 
The challenging aspect in this work is implementing a system that reliably detects identifying entities and substitute these recognized ones without compromising privacy. The following is our guiding example that will be developed throughout the article.

\begin{example}\label{exp_det}
Consider the following fictional sentence. It is typical of what can be present in a medical text document of a hospital.
"Mr. Durand born in Dijon, 40 years old, was admitted to the hospital from 12/02/2020 to February 26, 2020 following a road accident in Dijon".

\end{example}

To date, the most efficient methods in terms of Named Entity Recognition are those based on the attention concept, BERT~\cite{bert} for Bidirectional Encoder Representations from Transformers and its derivatives~\cite{le2020flaubert,clinicalbert,biobert}. To achieve detection scores where precision and recall are very high, they require previously labeled datasets for training. 
This kind of labeled dataset exists in English language~\cite{i2b2dataset,mimic2016}.
However, it is severely lacking in other languages, especially French.
This labeled learning dataset doesn't need to be perfectly coherent from a medical point of view. 
What is important is the format encountered and the context.
It then seems relevant to make use of an existing de-identification algorithm, even if imperfect, to provide this new dataset to be labeled afterward.

This article first shows how the utilization of a French dataset anonymized 
using a recent but not perfect de-identification algorithm followed by its manual annotation, allowed the implementation of a self-attention-based NER approach. The results obtained in terms of precision and recall exceed all existing approaches in French and are at the level of those in English.

After this NER phase, the following step is to substitute the detected sensitive entities.
This step is often neglected in research work because it is not considered relevant. This is indeed the case for entities such as phone numbers or email addresses that can be replaced by any random number or email address. The same is not true for dates or locations. 
Indeed, the chronology of medical events is essential in detecting correlations between them for example. Date substitution methods exist but are not satisfactory.
It has indeed been shown~\cite{tchouka:arxiv} that applying a uniform shift between dates allows guaranteeing the chronology but does not protect in any way a re-identification of the document. 
One could think of applying methods based on differential confidentiality~\cite{duchi2013local}, the only method to date providing a metric for the level of data leakage. Applied to temporal elements (date, age), it strongly protects privacy by making the original date indistinguishable from a published date.
However, it significantly degrades the usefulness of the data because of the magnitude of the interval in which the algebraic choice of the date to be published is made. 
This article shows how $d$-privacy brings a concrete answer to this problem of amplitude.

Finally, the location elements of textual medical records must be treated with great care. 
Randomly substituting a name of a city with another effectively protects privacy but this is done at the detriment of the medical context of the city. There was possibly radon in this one at the origin of cancers, and pollution at the origin of respiratory disorders. 
An approach based on geo-indistiguishability~\cite{andres2013geo} is not the most relevant since it only takes into account the geographical position and not the medical and/or statistical data associated with the city. 
We present in this paper an innovative approach based on $d$-privacy.

The result is a global approach to de-identification of medical documents dedicated to French textual documents but which could be generalized to any other language. This approach is first of all reliable in the detection of identifying entities. Based on differential privacy, substitution is robust to attacks by definition. Moreover, they are optimized to preserve data utility in the context of further processing by machine learning.

Our contributions in this paper can be summarized as follows:
\begin{enumerate}
    \item We provide a model identifying sensitive information according to HIPAA categories in clinical textual documents. This one manages to detect all the categories we want to detect, as well as to compete with the English detection models.
    \item We provide a robust surrogate generation approach based on advances in differential privacy that combines security and utility.
    \item An open-source implementation of the surrogate generation approaches proposed in this paper is available on GitHub\footnote{\url{https://github.com/healthinf/Surrogate-generation-Strategies-in-De-identification}}.
\end{enumerate}

This article is organized as follows.
The following section summarizes the state of the art regarding NER as well as the substitution of sensitive elements for de-identification purposes.
Section~\ref{sec:NER} shows how the NER task can be strengthened thanks to the construction of an annotated dataset on the one hand, and thanks to a deep learning-based model taking into account the context on the other hand.
Section~\ref{sec:surrogate} finally shows how to finely substitute temporal (age and date) and location entities without compromising the confidentiality of the data.
Finally, the last section presents a conclusion and future work. 

\section{{RELATED WORK}}\label{se:rel}

This section summarizes the state of the art of de-identification methods applied to the textual medical document.
The first section is dedicated to NER step whereas the second one focuses on the surrogate generation.

\subsection{Named Entity Recognition}
For the NER phase (English dataset), several works have experimented machine learning models such as SVM, Decision trees, or Condition Random Field (CRF)~\cite{CRFLafferty2001}. With the emergence of neural networks, researchers~\cite{Dernoncourt2016,Liu2017} have proposed the first neural network-based model. 
Recurrent neural networks (RNNs) of Dernoncourt et al~\cite{Dernoncourt2016} lead to $F_1$-scores of 97.85\% and 99.23\% on i2b2~\cite{i2b2dataset} and MIMIC~\cite{mimic2016} datasets respectively 
representing the state of the art in de-identification.
Some papers have obtained results almost as accurate as those of Dernoncourt by combining the machine learning method (CRF) and the neural recurrent network method (RNN). Following the recent advance of NLP with the emergence of transformers~\cite{vaswani2017attention} and BERT~\cite{bert} which are the state of the art in a contextualized text representation, it has been proven that the most accurate models for NER are those based on transformers. 
Among the abundant works in this field, we can cite~\cite{NERTransformersSoA} and~\cite{TransformerComparison}. Research on the de-identification of French medical documents is mainly done by C. Grouin~\cite{grouin2015possible} with a machine learning model (CRF reaching 80\% in F$_1$-score). 
A recent work~\cite{bourdois:hal-03241384} is dedicated to de-identification of French emergency medical records.
It is based on a twofold approach.
First, FlauBERT~\cite{le2020flaubert} assigns a label to documents which require de-identification. 
Next a combination of rules-based techniques and LSTM, via Flair~\cite{flair} is implemented. 
Unfortunately, there is no dataset like MIMIC or i2b2 in the French language. This forced the authors in~\cite{tchouka:arxiv} to combine the machine learning method (CRF done by C. Grouin) and the neural network method based on transformers on WikiNER dataset~\cite{wikiner} to integrate all the attributes to be detected.  This hybrid system reaches 94.7\% in F$_1$-score which serves as the baseline in this work.
\subsection{Surrogate Generation}
The complexity of the substitution phase~\cite{sweeney1996replacing} depends on the analysis of the documents. The most direct way is to delete the detected information or replace it with its entity name (Durand by NAME e.g.). This method protects privacy, but degrades the readability of the document and reduces the usefulness of the data. To preserve the structure of the document, several authors have tried other methods. 
The work of these papers~\cite{douglass2004computer,levine2003identification,uzuner2007evaluating,douglass2004computer,deleger2014preparing} has led to the following strategy: Names are replaced by a random name from a pre-established list, alphanumeric strings are replaced by a randomly generated string, and for dates, a uniform shift of days is performed while keeping the format. As for ages, they have been capped at 89 years, whereas locations are replaced randomly from a pre-established list.
The most used system in recent research is the system developed by Stubbs et al.~\cite{stubbs2015automated}.
This one combines the strategies of the previously described work. This system has been used to build the 2014 i2b2~\cite{kumar2015creation} dataset for example. The Stubbs method~\cite{stubbs2015automated} which consists in making a uniform shift of the dates of a document is easily attackable. Since the interval between the substituted dates remains unchanged, an attacker only needs to know one date in file to reconstruct the others. In~\cite{tchouka:arxiv} the authors have shown that the system proposed by Stubbs on dates and ages is easily attackable, thus compromising privacy in a medical context. Furthermore, for locations, this random method significantly degrades the level of information. The goal is to protect privacy while keeping as much information as possible. To do so, in~\cite{tchouka:arxiv} it was proposed to substitute dates and ages through the Local Differential Privacy (LDP)~\cite{duchi2013local} with the bounded Laplace mechanism~\cite{dwork2006calibrating} and to substitute locations by a geo-indistinguishability~\cite{andres2013geo} algorithm. The problem with LDP on dates or ages is that we cannot precisely control the noise added on two distant or close dates, which sometimes leads to inconsistencies in the document (e.g. the duration of a stay). The geo-indistinguishability method gives a coherent result but is not relevant in a medical context.

\section{STRENGTHENING NAMED ENTITY RECOGNITION}\label{sec:NER}

This section starts with our thread example. The first section starts with the motivation for the need to strengthen the NER stage. The second one shows how we obtained a new labeled medical dataset. Thirdly, the new machine learning-based NER approach is presented. Its evaluation on a medical dataset is finally presented in the fourth section.

\begin{figure}[ht!]
    \centering
    \includegraphics[width=0.99\columnwidth]{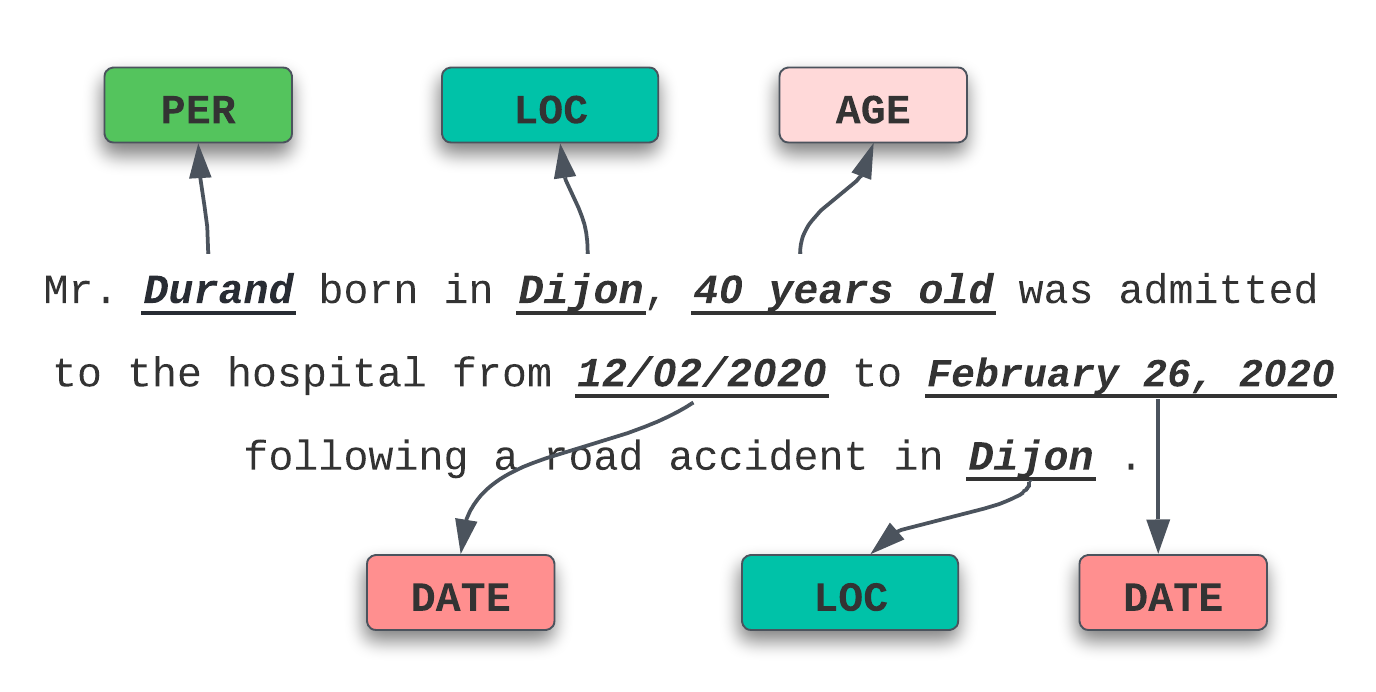}
    \caption{Perfect NER of PHI entities on thread example.}\label{fig:ner:xpl}
    \label{exple}
\end{figure}

\begin{example}
Figure~\ref{fig:ner:xpl} summarizes the result of a perfect NER process applied to the threaded example. 
HIPAA labels~\cite{cohen2018hipaa} with their descriptions are summarized in Table~\ref{tab:label}.

\end{example}

\begin{table}[ht!]
    \centering
    \scriptsize
    \begin{tabular}{|c|c|}
    \hline
    Label & Description \\
    \hline
    \textbf{PER} & All names of persons \\
    \hline
    \textbf{DATE} & All date sequences in all formats\\
    \hline
    \textbf{LOC} & All geographical locations and zip codes\\
    \hline
    \textbf{ORG} & Organizational entities\\
    \hline
    \textbf{AGE} & Ages\\
    \hline
    \textbf{TEL} & Phone Numbers\\
    \hline
    \textbf{REF} & All references related to individuals\\
    \hline
    \textbf{QID} & Any ID sequence\\
    \hline
    \end{tabular}
    \caption{Description of HIPAA labels.}
    \label{tab:label}
\end{table}

\subsection{Motivation and Global Overview}
Getting near-perfect scores in the NER is an absolute necessity for successful de-identification. Undetected sensitive information is a risk for re-identification of the document. The NER task is a problem-dependent task. This means that we often don't have the right dataset for our problem. The best results of NER in de-identification in the literature are the ones obtained by implementing English models (consistent and complete domain-specific English datasets). In French, such a dataset is rare and, to our knowledge, does not exist in the medical domain. It is, therefore, necessary to build a sufficient dataset adapted to the context of our application, \textit{i.e.} a medical corpus, which includes all the identifying attributes to detect. With such a dataset, we are then able to apply a Transformer based NER method which is the state of the art in NER. 

\subsection{Building a Labelled Dataset}
\begin{figure}[ht!]
    \centering
    \includegraphics[width=0.99\columnwidth]{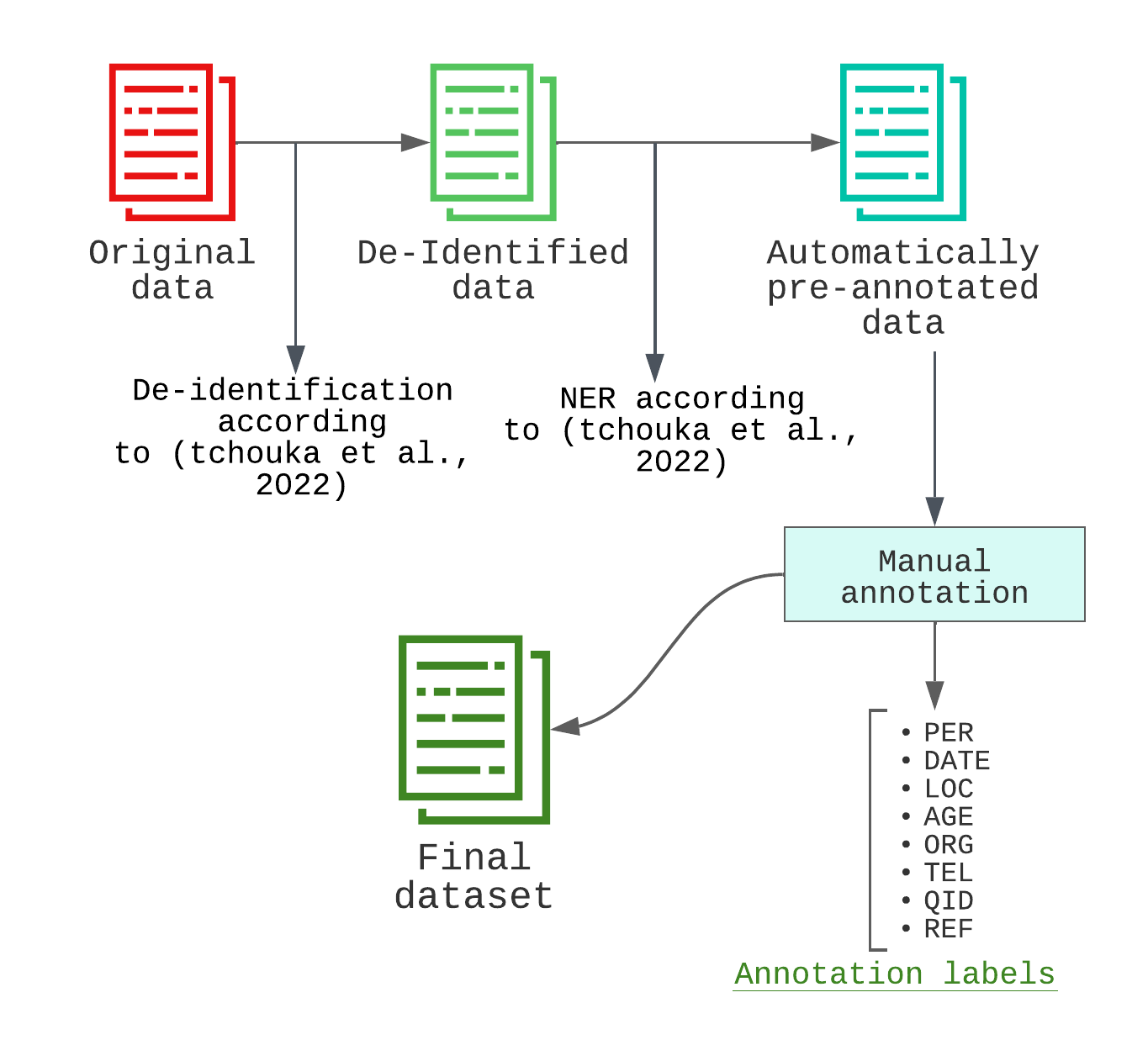}
    
    \caption{FrenchHospitalNER Dataset Construction.}
    \label{fig:dataset}
\end{figure}

As mentioned, the most difficult step is finding a dataset that is large enough for the implementation of an accurate model, that includes all the categories of sensitive information, and finally, that is adapted to the medical corpus. Such a dataset (in French) is not available at the moment, at least not accessible to everyone. 
In~\cite{tchouka:arxiv} the authors used WikiNER dataset which includes only a few tags with a very general vocabulary. 
This requires them to combine several methods so that all categories could be integrated into a  de-identification tool. 
In this current work, as part of our collaboration with a French public hospital, we have access (on-site) to a large set of unlabeled medical notes.
We propose that hospital members semi-manually annotate a subset of these notes.
As this dataset will be exported from the hospital afterward, we ask them to apply the existing de-identification method~\cite{tchouka:arxiv} on them to obtain new de-identified documents. 
Then we ask them to manually label all the de-identified documents.
To facilitate this manual task, the NER step of the same tool has been used. The process is illustrated in Figure~\ref{fig:dataset}. The obtained dataset, further denoted as FrenchHospitalNER is partially de-identified, according to the~\cite{tchouka:arxiv} approach, and labeled. This annotation step required 25 hours of work for one person (1 minute per file).

\subsection{Supervised Learning on a Dedicated Labelled Dataset}
This part starts by the presentation of the architecture of our model. Then we describe how supervised learning was implemented. Finally, an evaluation of our model is presented in parallel with existing de-identification models.

\subsubsection{Model Architecture: Transformer Based Approach}
Due to the the availability of FlauBERT which is a BERT-based pre-trained French models, such a transformer is easily accessible
This decision is strengthened by the fact that it has been shown in \cite{TransformerComparison}
that a language specifically dedicated to the French language model like FlauBERT, improves the results compared to multilingual BERT models.

\subsubsection{Finetuning Transformer Model}
\begin{figure*}[ht]
    \centering
    \includegraphics[width=0.9\linewidth]{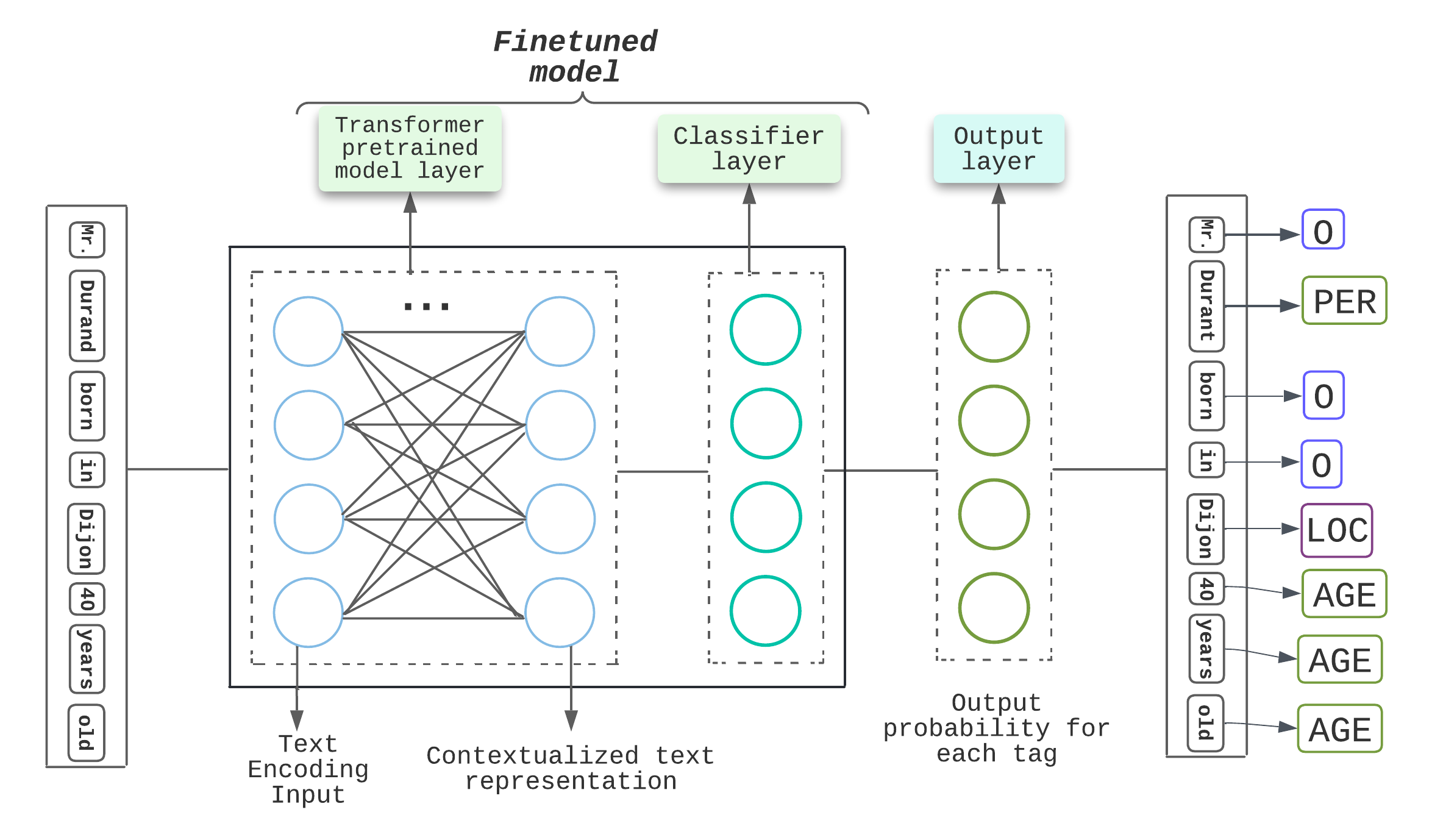}
    \caption{Deep Learning Model Architecture for NER.}
    \label{finetuning:architecture}
\end{figure*}

Starting from a pre-trained model such as FlauBERT, what is left is finetuning 
it on a smaller and more specialized dataset. 
Instead of starting from scratch to build our text classification or feature detection model, we will start from the pre-trained BERT and add a dense layer or a classification layer to build the model as described in  Figure~\ref{finetuning:architecture}.

\subsubsection{Training}

The learning process has been implemented with the previously described dataset using a deep learning-based NLP model.
\begin{itemize}
    \item the \textbf{Learning Rate} controls the size of the update steps along the gradient. Usually, a very small value is set ($10^{-4}$ in this work), so that the weights are less modified at each iteration, which avoids missing the optimal values of the error function
    \item the \textbf{Dropout} is a regularization technique for reducing overfitting in neural networks. It is set to $0.1$ in this work, which means that $10\%$ of selected neurons are ignored during training
    
    \item the \textbf{Training Batch Size} is the number of training samples to work through before the model’s internal parameters are updated.
    \item the \textbf{Maximum length} defines the maximum number of words in the sentences
    \item the \textbf{Number of epochs} is the number of complete passes through the training dataset.
\end{itemize}

The NER is a multi-class classification model (taking tags as classes). 
The CrossEntropy error function is well adapted for this task. As an optimizer (backpropagation function), we use the adamW\cite{loshchilov2017decoupled} algorithm which is one of the latest evolutions of optimizers and is proven to be better in neural network learning.

\begin{figure*}[ht]
    \centering
    \includegraphics[width=0.75\linewidth]{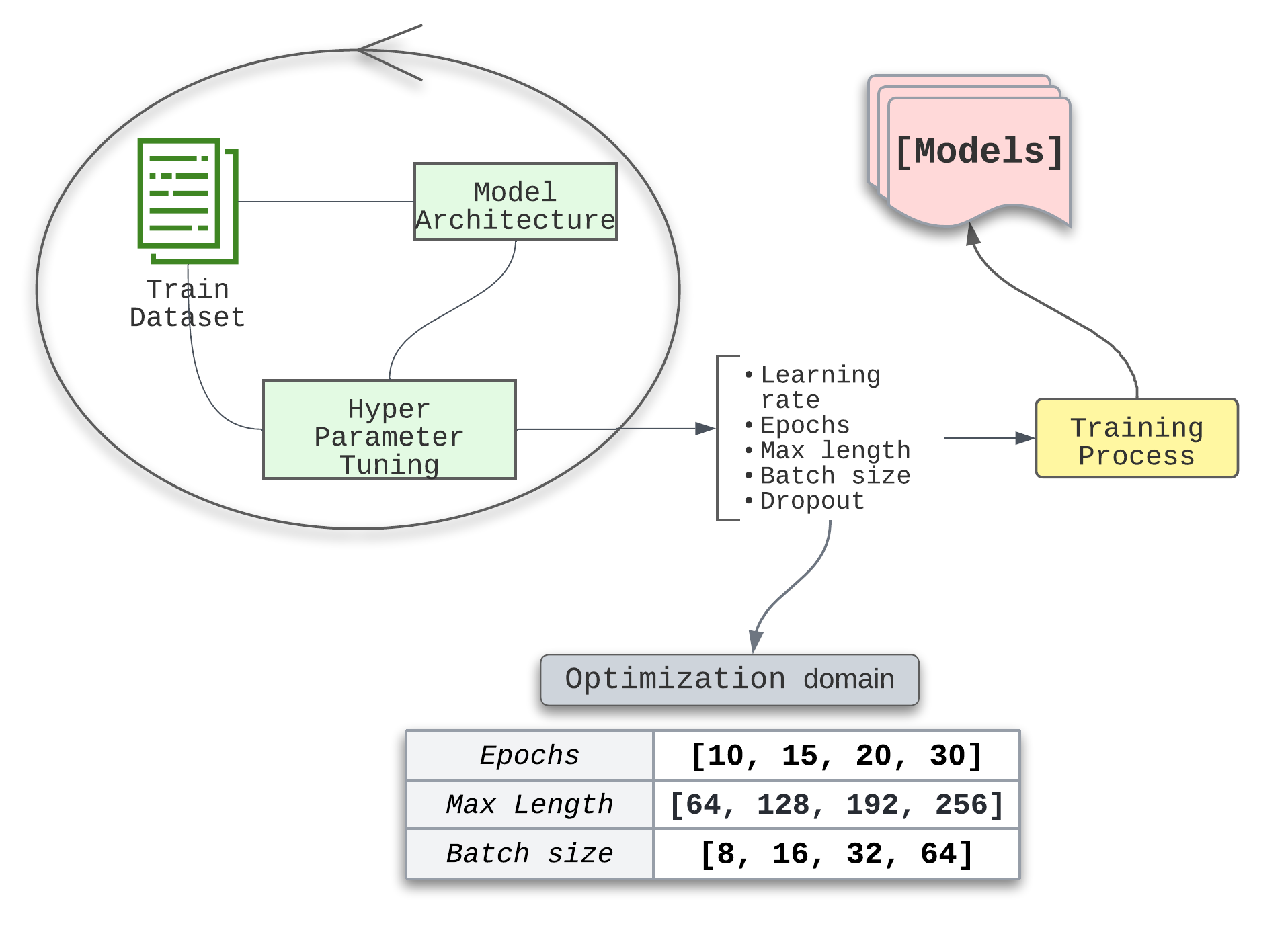}
    \caption{FineTuning Process.}
    \label{training}
\end{figure*}

In machine learning, the question remains: how to select the optimal values of the hyperparameters to obtain the most accurate results? There are several methods of hyper-parameter optimization such as Grid Search, Random Search, and model-based Bayesian method. Studies~\cite{bergstra2013making} on hyper-parameter optimization show that Bayesian methods give largely more accurate results. 
In this paper, for hyper-parameter optimization, we used the Tree-structured Parzen Estimator~\cite{bergstra2011algorithms} which is a classical Bayesian optimization algorithm sufficient for a classification model as in our case. 
We have experimented with different combinations of parameters according to the Tree-structured Parzen Estimator algorithm as illustrated in Figure~\ref{training}.

The optimization during training was performed on three hyper-parameters: the number of epochs, the maximum length, and the batch size.
At the end of the training, the model with the highest F$_1$ score is selected with the following hyper-parameters: \textit{number of epochs} = 20, \textit{maximum length} = 128, \textit{batch size} = 64. It is this model that is used in the Evaluation section below.

\subsection{Evaluation}

To evaluate our model we used the classical metrics Precision (P), 
Recall (R), and F$_1$-score.
To get a sense of the overall performance of the system, we use the micro-average of 
Accuracy of the labeling process is evaluated across precision, recall, and F$_1$-score metrics. 

\begin{table}[ht]
\centering
    \begin{scriptsize}
    \setlength{\tabcolsep}{1.5pt}
    \begin{tabular}{|l|c c c | c c c |c c c|}
    \hline
    Methods &
    \multicolumn{3}{c|}{\cite{tchouka:arxiv}}&
    \multicolumn{3}{c|}{{PROPOSAL}}& 
    \multicolumn{3}{c|}{\cite{Dernoncourt2016}}\\
    \hline
     Dataset & \multicolumn{6}{c|}{HNFC}&
     \multicolumn{3}{c|}{i2b2}\\
    \hline
     Metrics & P & R & F$_1$& P & R & F$_1$& P & R & F$_1$\\
     \hline
    PER &96.3 & {99.8} & 98 & 97.2 & 98.9 & {98} & {98.2} & 99.1 & 98.6\\
    \hline
    ORG & {41.1} & {57.3} & {47.8}& 90 & 51 & 65.6 & 92.9 & 71.4 & 80.7\\
    \hline
    LOC & 88.4 & {95.8} & {92}& 99.4 & 94.4 & {96.9} & 95.9 & 95.7 & 95.8\\
    \hline
    DATE & {97.7} & {86.7}& {91.9}& 99.2& 95.7 & {97.4}& 99 & 99.5 & 99.2\\
    \hline
    AGE & {91.5} & {66.9} & {77.3}& 98.2 & 91.8 & {95}& 98.9 & 97.6 & 98.2\\
    \hline
    TEL & {99.5} & {97.9} & {98.7}& 99.4 & 99.8 & {99.6}& 98.7 & 99.7 & 99.2\\
    \hline
    REF & & - & & 96.1& 79.5 & 87& & - & \\
    \hline
    QID & & - & & 77.2 & 32 & 45.3& 99.2 & 98.7 & 99\\
    \hline
    Overall & 94.6 &94.9 & 94.7 & {98.5} & 96.4 & {97.4}& 98.3 & 98.53 & 98.4\\
    \hline
    \end{tabular}
    \end{scriptsize}
    \caption{NER results on the evaluation dataset.} \label{tab:exp:results}
\end{table}

To be fair with~\cite{tchouka:arxiv}, we asked the HNFC hospital to evaluate the NER step of this approach on their HNFC-dataset. The results are detailed in Table~\ref{tab:exp:results}.
This proposal largely surpasses results obtained in~\cite{tchouka:arxiv} in several categories (DATE, AGE\ldots). This is due to the BERT-based layer which allows us to have a precise contextualization of the sequence.
Our low score on the organization level compared to the i2b2 model is explained by the fact that they are structured informally in the medical documents (abbreviation, isolated word\ldots). 
Increasing the dataset will help solve these types of problems and generally improve the scores in the different categories.

The next step is substituting the detected entities, as described in the next section.

\section{SURROGATE GENERATION STRATEGIES}\label{sec:surrogate}
The challenge here is to substitute personal information detected by NER 
with relevant surrogates regarding medical content whilst preserving privacy.  

As argued in~\cite{tchouka:arxiv,stubbs2015automated}, not all entities have the same level of criticality or importance.
A random strategy may be indeed chosen for instance for replacing names, 
phone numbers\ldots

Moreover, to avoid averaging attacks and for consistency in the document, memoization has been implemented as in~\cite{erlingsson2019amplification,arcolezi2022improving}. This consists in using the same substitute for given sensitive information in the document.

\begin{example}
In the thread example, Durand could be replaced by any name, Julien for instance.
\end{example}

In contrast, temporal and location data inherently carries information that is both medically important and highly identifiable.
In~\cite{tchouka:arxiv}, for temporal entities, the authors opted for 
local differential privacy  with bounds in time categories (recalled hereafter) 
to calibrate the added noise.
About geographical locations, geo-indistinguishability~\cite{andres2013geo} 
was retained as a direct mechanism to provide a location close to the original one and whose 
privacy leak is measured by $\epsilon-d$ privacy. 
These two approaches allow the aforementioned method to respect privacy. 
However, the relevance of substitutions in this specific context of medical data has some limits 
which will be detailed in each of the following two sections.

\subsection{Date \& Age: Substitution Strategy}

Beyond the fact that a date is identifying in a medical document, we can not afford to randomly substitute them. Providing an algorithm that respects privacy means accepting that only the patient is allowed to modify his or her data in such a way that given two sanitized data of two patients, it is difficult (from a probabilistic point of view) to reassign one to the first and the other to the second. Local Differential Privacy~\cite{duchi2013local} (LDP) formalizes the algorithm robustness and its definition is recalled hereafter.

\begin{definition}[$\epsilon$-local differential privacy]\label{def:ldp} A random mechanism ${\mathcal {A}}$ satisfies $\epsilon$-local differential privacy if, for any pair of input values $v_1, v_2 \in \text{Domaine}(\mathcal {A})$ and any possible output $y$ of ${\mathcal {A}}$ :

\begin{equation}
    \Pr[{\mathcal {A}}(v_1) = y]\leq e^{\epsilon }\cdot \Pr[{\mathcal {A}}(v_2) = y] \textrm{.}
\label{eq:ldp}
\end{equation}
\end{definition}

LDP mechanisms~\cite{holohan2017optimal} are tuned with respect to the data types  they handle (real, integer, \ldots),
and to their usefulness. 
Here, the focus is on temporal data. As each of them can be seen as a number (number of days elapsed between the date to be cleaned and the current date), in~\cite{tchouka:arxiv} the authors have focused on the Laplacian mechanism recalled below.   

\begin{definition}[Laplacian mechanism in an interval of amplitude $\Delta$]\label{def:lapl}
{\sloppy In the Laplacian mechanism, a numerical value $v$ is sanitized into a numerical value
${\mathcal {M}}_{\mathrm {Lap} }(v,\Delta,\epsilon)$
with} 
\begin{equation}
    {\mathcal {M}}_{\mathrm {Lap} }(v,\Delta,\epsilon )=v+\mathrm {Lap} \left(\frac{\Delta}{\epsilon }\right)
\label{eq:lapl}\end{equation}
\end{definition}
\noindent where $\mathrm {Lap} \left(\frac{\Delta}{\epsilon }\right)$ is the Laplace distribution centered in 0 and whose scale parameter is $\frac{\Delta}{\epsilon }$.

In~\cite{tchouka:arxiv}, the authors reached the conclusion that segmenting a set of dates into 3 categories (less than 2 months, less than 2 years, more than two years) was necessary to minimize $\Delta$, \textit{i.e.} the introduced noise. 
Indeed, within these intervals (of range $\Delta$), the generated dates did not allow us to infer what their preimages were. However, in the larger category, the introduced noise is still too important since it is necessary to make indistinguishable the cleaning of two dates like 3 years and 80 years. 

It is thus necessary to further segment the space much more or, equivalently, allow distinguishing certain dates from others. 
Two dates that are initially far apart should not necessarily be made identical by a differential privacy mechanism.
The underlying idea is therefore privacy depending on the distance between the values of the elements to be protected. We find here the notion of $\epsilon.d$-privacy~\cite{alvim2018metric} recalled below.
\begin{definition}[$\epsilon.d$-privacy]\label{def:edp} 
A randomized algorithm ${(\mathcal {A}}$ satisfies the $epsilon.d$-privacy if, 
for any possible output $y$ of ${\mathcal {A}}$
and for any pair of input values $v_1, v_2 \in \text{Domain}(\mathcal {A})$,
domain with a metric $d$.

\begin{equation}
    \Pr[{\mathcal {A}}(v_1) = y]\leq e^{\epsilon.d(v_1,v_2)} \cdot \Pr[{\mathcal {A}}(v_2) = y] \textrm{.}
\label{eq:edp}
\end{equation}
\end{definition}

Intuitively, the $\epsilon.d$-privacy protects the precision of the secret: if we add a metric in the date space, it allows us to distinguish between an old date (of birth for example) and a recent date (of operation last week). On the other hand, it guarantees that two very recent dates ($v_1$ and $v_2$) at a very small distance will generate the same output $y$ with a very high probability.

Without going into further details, this version of $\epsilon.d$-privacy generalizes both the 
$\epsilon$-local differential privacy and $\epsilon$-differential privacy and has the same properties in terms of composition and post-processing~\cite{fernandes2021differential}. 

The question then arises of implementing a mechanism that guarantees this $\epsilon.d$-privacy property for temporal events.

In~\cite{tchouka:arxiv}, to respect the chronology of events each temporal event $d$ (a date, an age,\ldots) is converted into a duration $v$ in days between the current date and $d$.
The input domain is thus $\mathbb{R}^{+}$ with the absolute value as distance.
Easy to implement, if it were applied as is here, it would be detrimental to privacy. 
Indeed, in the context of $\epsilon$-LDP, the $\Delta$-amplitude of this mechanism would be equal to 1 day instead of the amplitude of each category ($100\times 365$ days for the largest category). A precise date such as birth or intervention will probably be modified. On the other hand, 
an age of a few decades will very probably not be modified, which is not satisfactory.

Moreover, in a medical document which contains "10 years ago", what is actually meant is "about 10 years ago", and definitively not "the same day, 10 years before".
This approximation is also found when temporal events are expressed in months or weeks.
The metric that we will consider will be unit dependent. It will be in years (in months, weeks, and days respectively) for events expressed in years (in months, weeks, and days respectively).
With this adaptation, an age ( in years) will probably be modified by a few years, for example. 

The Laplacian mechanism (recalled in definition~\ref{def:lapl}) adds noise following a Laplace-centered distribution of parameter $\epsilon^{-1}$. It is not difficult to demonstrate~\cite{fernandes2021differential} that this mechanism has the $\epsilon.d$-privacy property given by the equation~\eqref{eq:edp}.

Notice that, as in a classical differential privacy approach, 
the privacy global $\epsilon$ budget is shared between all the elements to be substituted. 
This sharing here can be uniform or not. Without any a priori, we consider that it is here.

\begin{example}
    In our thread example, there are 2 detected dates (expressed in days), 1 age expressed in years, and 1 location (2 times duplicated), \textit{i.e.}
    4 elements to substitute.
    Each element will consume $\frac{\epsilon}{4}$ of the privacy budget, the last quarter is dedicated to sanitizing location. 
    The date substitution process for this example is detailed in Figure~\ref{date_process}.
    Thus, \textbf{40} years becomes 37 years, \textbf{02/12/2020} and \textbf{February 26, 2020} respectively lead to  02/20/2020 and March 01, 2020.

    \begin{figure}[ht!]
    \centering
    \includegraphics[width=0.99\columnwidth]{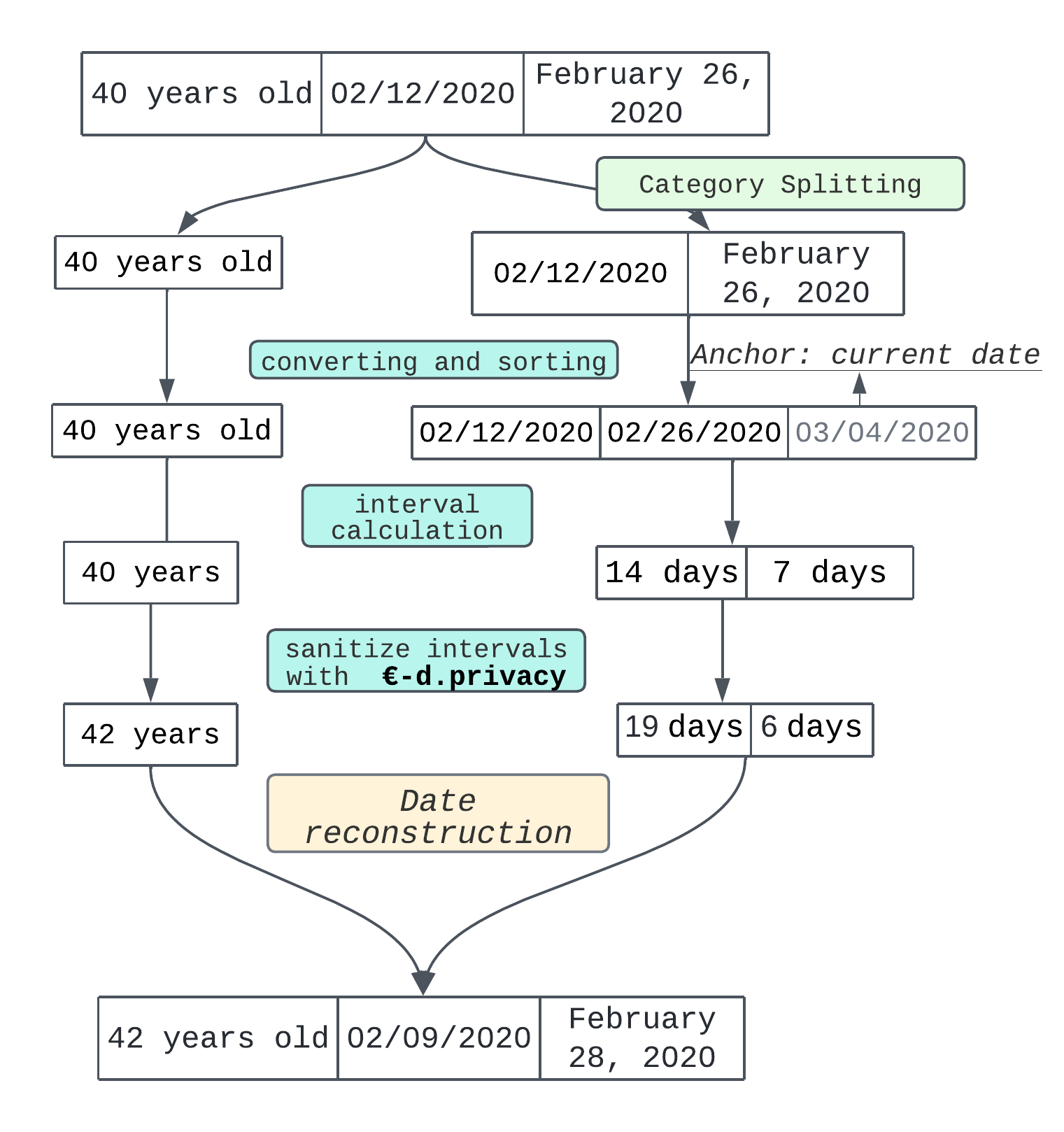}
    \caption{Example of date substitution process on the thread example.}
    \label{date_process}
    \end{figure}
\end{example}

\subsection{Geographic Locations: Substitution Strategy}
Geo-indistinguishability~\cite{andres2013geo} has been accepted de facto as the gold standard to preserve location privacy~\cite{xiao2015protecting,fawaz2014location,bordenabe2014optimal}. 
This mechanism instantiates $\epsilon.d$-privacy (as recalled in definition~\ref{def:edp}) in the context of locations  
which are $(x,y)$ coordinates inside $\mathbb{R}^2$. 
In the de-identification context, the authors of~\cite{tchouka:arxiv}
use geo-indistinguishability to randomly add noise to the coordinates $(x,y)$ of the location to be sanitized leading to the new tuple $(x', y')$. Thus, they re-associate the location which is the 
closest one to this new tuple. 
This method effectively protects privacy and provides a consistent substitute in the document. 
However, we argue that it does not effectively answer the question of the document's utility in a medical context. 
Indeed, two places relatively close to each other geographically can be far from each other from a health point of view.
Our motivation here is to have a system that integrates not only the distance but also some health criteria that can impact the health of the population.

\begin{figure*}[!ht]
    \centering
    \includegraphics[width=0.80\linewidth]{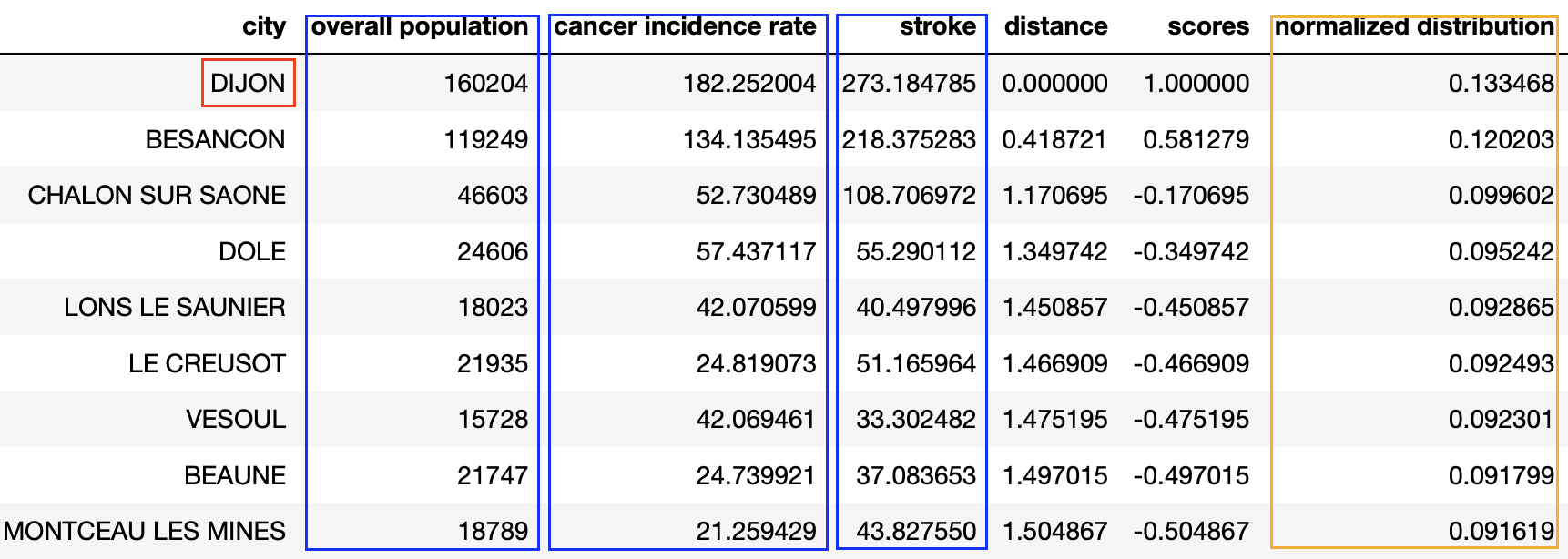}
    \caption{Example of exponential mechanism applied on sanitizing Dijon city.}
    \label{dijon-mech}
\end{figure*}

In this substitution of locations, it seems desirable to be able to choose randomly among locations that are close not only in a geographical sense but also in a statistical (e.g., number of inhabitants) and medical (e.g., the incidence rate of all cancers, the number of strokes, air pollution, radon level\ldots) sense.
Everything depends on the fact that we can express a distance between locations that would integrate statistical and medical characteristics. In practice, many institutional websites freely offer this local information.  
Figure~\ref{dijon-mech}, in its blue framed part, gives an extract for some cities of the Bourgogne Franche-comté, a region of France.
With such features for each location, it is not hard to compute the distance between them (for instance the euclidean one) and to apply any LDP mechanism capable of capturing this distance.

For instance, here, let us consider a public database of $N$ locations where each location $i$ 
is a vector 
$(x_i, y_i, c^1_i, \dots, c^n_i)$ where $(x_i, y_i)$ is the geographical location and $(c^1_i, \dots, c^n_i)$ the features. 
Hereafter, we consider all features to be normalized, \textit{i.e.} in $[0,1]$.
Let $d_{ji}$ be the vector of feature differences between locations $j$ and  $i$.

Let $v_j = [(1,d_{j1}),(2,d_{j2}), \dots(j, 0),\dots, (N,d_{jN})]$ 
be the sequence of all distances between $j$ and others. 
In a practical situation, this sequence can be reduced to the location distances $(i,d_{ji})$ s.t. both the geographical distance between $i$ and $j$ is lower than a given threshold and to the $k$ smallest values of the distances,
and where values are sorted in ascending order according to $d$.
$v'_j$ is the result and constitutes the possible substitutes of the city $j$. 
This leads to $v'_j = [(i_1,d_{ji_1}), \dots(i_k,d_{ji_k})]$, with $(i_1,d_{ji_1})=(j,0)$ since the smallest distance is 0 between $j$ and $j$. The score function $U$ may be defined by 
$U(j,i)= 1 - d_{ji}$ for each $i \in\{i_1, \dots, i_k \}$ and $-\infty$ elsewhere. 
This function is public and is not based on any private data.

The probability distribution function is thus as follows:
\begin{equation}
    P_j=[a.e^{\epsilon U(j,i_1)},\dots,a.e^{\epsilon U(j,i_k)},0,\dots,0] 
    \label{eq:Pj}
\end{equation}
where $a=\left(\sum_{i=1}^k e^{\epsilon U(j,i_1} \right)^{-1}$ is the normalization factor.
Notice this mechanism is an adaptation of the centralized exponential mechanism with public data, \textit{i.e.}, without sensitivity.
Cities can thus be sanitized according to the mechanism given in algorithm~\ref{alg_loc:cap}.
This mechanism is based on a distance? The next section shows it verifies   
$\epsilon.d$-privacy.


\begin{algorithm}[t]
\caption{Local exponential mechanism applied to the  city $j$ }{\label{alg_loc:cap}}
\scriptsize
\begin{algorithmic}
\item
\begin{itemize}[label=\ding{212}]
\item Let the probability distribution $P_j$
defined as in ~\eqref{eq:Pj}
\item $Y_j = [y_1, \dots, y_k]$ the $k$ possible output cities 
\item the substitute $l$ of the city $j$ :
\[
    l = Random[Y_j]_{P_j}
\]

with $Random[Y_j]_{P_j}$ a random draw according to the distribution $P_j$
\end{itemize}
\end{algorithmic}
\end{algorithm}

\begin{property}

The mechanism defined in Algorithm~\ref{alg_loc:cap} verifies   
$\epsilon.d$-privacy.
\end{property}
\begin{proof}
According to the definition \ref{def:edp}, for any $y$ whose 
probability distribution definition is not null we successively have

$\begin{array}{ll}
\frac{\Pr[{\mathcal {A}}(v_1) = y]}{Pr[{\mathcal {A}}(v_2) = y]} & =  \frac{a e^{\epsilon U(v_1,y)}}{a e^{\epsilon U(v_2,y)}} = \frac{e^{\epsilon (1 - d(v1,y))}}{e^{\epsilon (1-d(v2,y))}} \\
        & = e^{\epsilon (d(v_2,y) - d(v_1,y))}
    \leq e^{\epsilon.d(v_1,v_2)}
    \end{array}
$

\noindent according to Euclidean distance property.
\end{proof}

\begin{example}
Using our example with the location "Dijon". Considering the features: overall population, cancer incidence rate, and strokes, shown in blue in Figure~\ref{dijon-mech}. 
The columns ('distance' \& 'scores') represent respectively the vector distance (Euclidean distance with normalized features) and the results of the score function defined in Algorithm~\ref{alg_loc:cap}, from Dijon to $k=10$ 'nearby' cities (according to features). After applying the probability distribution function previously detailed, we obtain the normalized distribution illustrated in orange in Figure~\ref{dijon-mech}. The random draw thus follows this distribution.

According to the memoization, all occurrences of the location \textbf{Dijon} can be replaced by Besançon.

\end{example}

The final result of the substitution step would be: \textit{"Mr. Julien born in Besançon, 37 years old, was admitted to the hospital from 02/20/2020 to March 01, 2020 following a road accident in Besançon"}
\section{{CONCLUSION}}
\label{sec:conclusion}

This paper detailed a complete accurate differentially private de-identification method.
Regarding the NER step, an existing comprehensive but flawed de-identification approach was taken 
to internally build a new and substantial medical dataset that was then labeled by hand.
Using this new labeled and large dataset, deep learning was implemented taking into account the context. NER results we obtained in French are equivalent to the most accurate results in the English language, filling the gap between de-identification methods in these two languages.
Regarding substitutions of sensitive data, we pointed out the limitations of existing approaches, especially for temporal and location data.
We believe we have provided the most privacy-friendly method to date and location (since it is based on differential privacy) that retains sufficient medical information for further processing.    
\noindent
For the NER part, our future works will focus on the use of multilingual models
such as XLM-RoBERTa and their ability to enable zero-shot cross-lingual transfer. The idea here is to fine-tune a such model on an English NER medical dataset and to study its behavior on a French evaluation dataset.
\noindent
With the same goal of analyzing medical documents while preserving privacy, in addition to the anonymization method detailed in this paper, we will try the methods of perturbing the training dataset in the word embedding vocabulary (BERT-based model) by metric-based differential privacy as in~\cite{feyisetan2020privacy,zhao2022survey}

\bibliographystyle{apalike}
{\small
\bibliography{reference}}

\end{document}